\documentclass[12pt]{article}
\usepackage{epsf, cite, amssymb}
\usepackage{epsfig}

\makeatletter
\renewcommand\section{\@startsection {section}{1}{\z@}%
                                   {-3.5ex \@plus -1ex \@minus -.2ex}
                                   {2.3ex \@plus.2ex}%
                                   {\normalfont\large\bfseries}}
\renewcommand\subsection{\@startsection{subsection}{2}{\z@}%
                                     {-3.25ex\@plus -1ex \@minus -.2ex}%
                                     {1.5ex \@plus .2ex}%
                                     {\normalfont\bfseries}}
\makeatother

\newcommand{\bea}{\begin{eqnarray}}
\newcommand{\eea}{\end{eqnarray}}
\newcommand{\be}{\begin{equation}}
\newcommand{\ee}{\end{equation}}

\newcommand{\noi}{\noindent}

\begin{document}
\begin{titlepage}

\begin{center}

\hfill ITFA-07-29

\vskip 2 cm {\Large \bf AdS$_3$ Partition Functions Reconstructed} \vskip
1.25 cm {Jan Manschot}\\
{\vskip 0.5cm  Institute for Theoretical Physics\\ University of Amsterdam\\
Valckenierstraat 65\\
1018 XE Amsterdam\\
The Netherlands\\}

\end{center}

\vskip 2 cm

\begin{abstract}
\baselineskip=18pt

\noi For pure gravity in AdS$_3$, Witten has given a recipe for the
construction of holomorphically factorizable partition functions of pure
gravity theories with central charge $c=24k$. The partition function
was found to be a polynomial in the modular invariant $j$-function. We
show that the partition function can be obtained instead as a modular
sum which has a more physical interpretation as a sum over
geometries. We express both the $j$-function and its derivative in
terms of such a sum.  
\end{abstract}

\end{titlepage}

\pagestyle{plain}
\baselineskip=19pt

\section{Introduction}
Partition functions of gravity in three dimensions with a negative
cosmological constant are strongly constrained by modular
invariance. Witten recently used this constraint in
Ref. \cite{Witten:2007kt} to construct partition functions 
for pure gravity theories which allow a holomorphically factorized
partition function. The constraint of modular invariance
 is strong enough to determine a conformal field theory partition function completely from the lowest terms of its
Laurent expansion, as demonstrated earlier
in\cite{Dijkgraaf:2000fq}. Ref. \cite{Witten:2007kt} uses these lowest
terms to express the partition function as a polynomial in the
modular invariant $j(\tau)$. However, the partition function as a polynomial in 
$j(\tau)$ does not display any apparent connection to the gravity
path integral. 

We would like to emphasize in this note that each of the partition
functions obtained in Ref. \cite{Witten:2007kt} can be written as a sum over a coset
of the modular group. This sum has a clear interpretation as a sum
over geometries. Ref. \cite{Dijkgraaf:2000fq} first wrote the partition
function in this way for the D1-D5 system, where the expansion was
given the name ``Farey tail.'' In the past year there have been several
applications of these techniques to $\mathcal{N}=2$ supersymmetric black holes in four
dimensions\cite{deBoer:2006vg, Gaiotto:2006wm, Denef:2007vg,
  kraus-2007-0701}.  

\section{Gravity action}
Ref. \cite{Witten:2007kt} is mainly concerned with pure gravity without a
gravitational Chern-Simons term. This gives rise to a partition
function with a holomorphic as well as an anti-holomorphic dependence. A
subclass of partition functions are those which can be holomorphically
factorized. The holomorphic or anti-holomorphic part can be studied
independently in those situations. 

In this note, we choose to restrict to theories whose partition
functions are holomorphic by adding an appropriate Chern-Simons term to
the standard Einstein-Hilbert action. The action is the common
Einstein-Hilbert action plus the gravitational Chern-Simons term (in
Euclidean signature)\cite{kraus-2007-0701} 

\be
S_\mathrm{grav}=\frac{1}{16\pi G}\int d^3 x
\sqrt{g}\left(R-\frac{2}{l^2}\right) + \frac{k'}{4\pi}\int d^3 x \Omega_3(\omega),
\ee

\noindent where $\Omega_3(\omega)$ is the holomorphic Chern-Simons
form,

\be
\Omega_3(\omega)=\omega\wedge d\omega +\omega\wedge \omega \wedge \omega.
\ee 

\noindent We introduce a gauge field $A_L=\omega-^*e/l$ and
$A_R=\omega+^*e/l$. The action in terms of these variables is 

\be
S_\mathrm{grav}=\frac{k_L}{4\pi}\int A_L\wedge dA_L+\frac{2}{3}A_L\wedge A_L\wedge A_L-\frac{k_R}{4\pi}\int A_R\wedge dA_R+\frac{2}{3}A_R\wedge A_R\wedge A_R,
\ee

\noindent with $k_L=\frac{l}{16G}+\frac{k'}{2}$ and
$k_R=\frac{l}{16G}-\frac{k'}{2}$. Our aim is to study a holomorphic
theory, so we take $k_R=0$, which gives
$k_L=\frac{l}{8G}=k$. Quantum mechanical consistency requires $k$ to
be an integer.

Gravity in three dimensions has no local degrees of freedom. Different
geometries are determined by globally different identifications. The
path integral therefore reduces to a sum over these
identifications. We can determine the action for
different geometries. The action of thermal AdS$_3$ with $k_R$ equal to 0 is\cite{Maldacena:1998bw}

\be
S=2\pi i k \tau.
\ee

\noindent The action of the BTZ black hole is 

\be
S=-\frac{2\pi i k}{\tau}.
\ee

\noi The action of the BTZ black hole and thermal AdS$_3$ are related
by the transformation $\tau\to -\frac{1}{\tau}$, which is a generator
of $SL(2,\mathbb{Z})$. Ref. \cite{Dijkgraaf:2000fq} shows that the
geometries of AdS$_3$ are in one-to-one correspondence with the coset
$\Gamma_\infty\backslash SL(2,\mathbb{Z})$, where $\Gamma_\infty$ is
the group of
``translations'' given by $\left(\begin{array}{cc}1 & r \\ 0 & 1
\end{array}\right)$. Any element $\left(\begin{array}{cc}a & b \\ c & d
\end{array}\right)\in  \Gamma_\infty\backslash SL(2,\mathbb{Z})$ is
determined by a choice of two relatively prime integers $c$ and $d$. This set of different geometries comes about as 
different choices of the primitive contractible cycle when Euclidean AdS$_3$ is
viewed as a filled torus\cite{Dijkgraaf:2000fq}. The action of these other geometries is
given by $2\pi i\left(\frac{a\tau +b }{c\tau +d}\right)$. The gravity partition
function is now given by

\be
\label{eq:gravpart}
Z_{k}(\tau)=\sum_{\mathrm{geometries}}e^{-S}=\sum_{\Gamma_\infty
  \backslash SL(2,\mathbb{Z})}e^{-2\pi i k \frac{a\tau+b}{c\tau+d}}M(c\tau+d),
\ee

\noi where $M(c\tau+d)$ is some measure factor. Such sums over the
modular group are known as Poincar\'e series in the mathematical
literature. Modding out the translations $\Gamma_\infty$ from
$SL(2,\mathbb{Z})$ is necessary for convergence in such sums. 

\section{Conformal field theory partition function}
The AdS/CFT correspondence relates the degrees of freedom in the bulk
of AdS space including gravity to a conformal field theory on the
boundary. Ref. \cite{Witten:2007kt} argues that for pure gravity the whole
partition function can be constructed from knowledge of the ground state. The
ground state energy is given by $-k=-c/24$, where $c$ is the left
moving central charge. States other than the vacuum and its descendants must
be related to black holes, because gravity in three dimensions has 
no local degrees of freedom. Primary states other than the vacuum do
not have negative energy because black holes with negative mass
do not exist. Therefore, all polar terms (i.e. $q^{-n}$, $n>0$) in
the partition function are the vacuum and its descendants. The vacuum
$\left|0\right>$ is primary and $SL(2,\mathbb{R})$ invariant and is thus
annihilated by $L_{n}$ for $n\geq -1$. Acting with creation operators
$L_{-n}$, $n \geq 2$  generates a tower of states with partition function

\be
\label{eq:trun}
Z_{\mathrm{subset},k}(\tau)=q^{-k}\prod_{n=2}^\infty \frac{1}{(1-q^n)},
\ee

\noi We gave this partition function the subscript ``$\mathrm{subset}$''
because it represents only a subset of the total number of 
states in the theory. A direct way to see this is that the partition
function is not modular invariant. Ref. \cite{Witten:2007kt} constructs a
modular invariant partition function with the required polar behavior
with the use of $J(\tau)=j(\tau)-744$, the unique modular invariant with a polar
term $q^{-1}$ and vanishing $q^0$ term. $j(\tau)$ is given by

\be
j(\tau)=\frac{1728 E_4(\tau)^3}{\Delta(\tau)}=q^{-1}+744 +
\sum_{n=1}^\infty c(n) q^n,
\ee

\noi where $\Delta=\eta(\tau)^{24}$, and $E_4(\tau)$ is the familiar
 Eisenstein series of weight 4. The partition function for $k=1$ is
 equal to $Z_1(\tau)=J(\tau)$. The partition functions for larger values of $k$
 become polynomials in $J(\tau)$.

The exact Fourier coefficients of $j(\tau)$ can be determined with the
circle method introduced by Rademacher.\footnote[1]{Ref. \cite{apostol}
  contains a clear exposition of the circle method
  applied to the Dedekind $\eta$-function.} This method to determine
Fourier coefficients was originally obtained for modular forms of
negative weight and with a polar part. It also turned out to be
very useful for the determination of the Fourier
coefficients of $j(\tau)$ which has weight 0. The coefficients $c(n)$
are given by the infinite sum\cite{petersson:1932,rademacher:1938}

\be
\label{eq:radseries}
c(n)=\frac{2\pi}{\sqrt{n}}\sum_{m=1}^\infty
\frac{K_m(n)}{m}I_1\left(\frac{4\pi\sqrt{n} }{m}\right),
\ee

\noi where $K_m(n)$ is the Kloosterman sum 

\be 
K_m(n)=\sum_{ d\in (\mathbb{Z}/m\mathbb{Z})^*}\exp\left(\frac{2\pi i
  (nd + \bar d)}{m} \right), \qquad d\bar d=-1\,\, \mathrm{mod}\,\, m,
\ee

\noi and $I_\nu\left( z \right)$ is the Bessel function defined by 

\be
I_\nu(z)=\frac{\left(\frac{1}{2}z\right)^\nu}{2\pi i}
\int_{c-i\infty}^{c+i\infty}t^{-\nu-1}e^{t+\frac{z^2}{4t}}dt, \qquad
\left(c>0, \mathrm{Re}(\nu)>0\right).
\ee

\section{Partition function as a sum over geometries}
We would like to relate the conformal field theory partition
function to the gravity partition function (\ref{eq:gravpart})
in a spirit similar to \cite{Dijkgraaf:2000fq}. There exist in fact two sums
over integers $c$ and $d$ that are relatively prime $((c,d)=1)$, which are related to the modular
invariant $J(\tau)$. We will comment on both.
 
Refs. \cite{rademacher:1939, knopp:1990} give for $J(\tau)$ 

\be
\label{eq:modsum1}
J(\tau)=-12+\lim_{K\to\infty} \frac{1}{2}\sum_{|c|\leq
  K}\sum_{{|d|\leq K}\atop
{(c,d)=1}} \exp 2\pi 
  i\left(-\frac{a\tau+b}{c\tau+d}\right)-s(c,d), \qquad ad-bc=1,
\ee

\noi where $s(c,d)$ is defined as $\exp 2\pi i
\left(-\frac{a}{c}\right)$ when $c\neq 0$ and otherwise $0$. The
subtraction of $s(c,d)$ is necessary for convergence. The order of
summation over $c$ and $d$ is important in this case. For every
integer pair $(c,d)$, $a$ and $b$ are chosen to satisfy $ad-bc=1$. Two differences
with the gravity partition function, Eq. (\ref{eq:gravpart}), are the
subtraction of $s(c,d)$ and the dependence on the order of
summation of $c$ and $d$. The sum over $(c,d)$ can still be
interpreted as different choices of the primitive contractible
cycle as in Eq.(\ref{eq:gravpart}), but the subtraction of $s(c,d)$
might be harder to interpret from the gravity point of view. The $q^0$
term is not determined by the modular sum, because it is itself a
modular form of weight 0. Similar sums are known for other modular
forms with negative (integer) weight, although they might transform
with a shift\cite{knopp:1990}.  

A way to cure the discrepancies between the gravity partition function
and the sum in (\ref{eq:modsum1}) is to consider the so-called Farey
transform of the partition function. The Farey transform of the weight
zero partition function $Z_k(\tau)$ is simply the derivative
$DZ_k(\tau)$, where we defined the differential operator 
$$D=\frac{1}{2\pi i}\frac{d}{d\tau}.$$  

\noi Thus the Farey transform of $J(\tau)$ is
$DJ(\tau)$. The inverse transform gives back $J(\tau)$ up to the
constant term. Calculation of the Fourier coefficients of the relevant
Poincar\'e series\cite{Sarnak:1990} shows that these are equal to
those of the Farey transformed partition
function. Ref. \cite{petersson:1932} gives the Poincar\'e series for 
$DJ(\tau)$ as

\be
\label{eq:farey}
DJ(\tau)=-\frac{1}{2}\sum_{\Gamma_\infty\backslash SL(2,\mathbb{Z})}\frac{\exp 2\pi i\left(- \frac{a\tau +b}{c\tau +d}\right)}{(c\tau+d)^2}.
\ee

\noi The fact that $DJ(\tau)$ is a weight 2 modular form makes a
convergent series possible irrespective of the order of the summation
over $c$ and $d$. The sum is now much more reminiscent of Eq. 
(\ref{eq:gravpart}), giving it a natural interpretation as a sum over
geometries. The measure factor introduced in Eq. (\ref{eq:gravpart}) is determined to be
$M(c\tau+d)=-\frac{1}{2}(c\tau+d)^{-2}$. In this sense the sum gives a
physical explanation of the modular invariance and shows moreover how
the complete partition function is obtained from knowledge of the
polar part of the partition function.   

The construction of the transformed partition function for larger values of $k$
is straightforward now that we know it for $k=1$. Take the derivative of
 $Z_{\mathrm{subset},k}(\tau)$ and perform a Laurent expansion up to the
constant term (similarly to \cite{Witten:2007kt}): 

\be
D \tilde Z_k(\tau)=\sum_{-k\leq r<0} a(r)q^{r}.
\ee

\noi Then the derivative of the total partition function is given by

\be
\label{eq:fareygen}
D Z(\tau)=-\frac{1}{2}\sum_{-k\leq r<0} \sum_{(c,d)=1}a(r)\frac{\exp 2\pi i\left(r\frac{a\tau +b}{c\tau +d}\right)}{(c\tau+d)^2}.
\ee

Partition functions for larger values of $k$ can also be written
as sums analogous to Eq. (\ref{eq:modsum1}). The resulting series have
with Eq. (\ref{eq:fareygen}) in common, that they are both  
modular sums of polar terms. The states corresponding to the polar
terms have a physical interpretation as states which are not sufficiently 
massive to form black holes. The mass of a black hole is given in the
holomorphic case by $M=\frac{1}{l}(L_0-\frac{c_L}{24})$ and a black
hole is only formed when $M \geq 0$. Note that in
principle, terms $q^r$  ($r>0$) could be included in the sum. The sum
over these terms would vanish since cusp forms do not exist for weight
0 and 2. 

We have given arguments to interpret  holomorphic partition
functions as sums over geometries. However, Ref. \cite{Witten:2007kt} does not
consider holomorphic partition functions but holomorphic
factorizable partition functions. An example of such a partition
function is $Z_1(\tau,\bar \tau)=|J(\tau)|^2$. Application of the sums
in Eqs. (\ref{eq:modsum1}) or (\ref{eq:farey}) leads to a sum over
$(c,d)$ and $(\tilde c, \tilde d)$, one pair for the holomorphic side and
one for the anti-holomorphic side.\footnote{I would like to thank E. Witten and the referee for
  bringing this point to my attention.}
 Only the terms with  $(c,d)=(\tilde c, \tilde d)$ correspond to
 classical geometries. This raises the puzzle that  holomorphically
 factorizable partition functions require states  which are difficult
 to interpret classically.   

\section{Conclusion}
We have considered the question of how to construct
holomorphic partition functions of pure gravity in AdS$_3$ for given
central charge. We emphasized the fact that the partition function can
be written as a sum over $\Gamma_\infty\backslash
SL(2,\mathbb{Z})$. We presented two such sums: one for $J(\tau)$, and one
for its Farey transform $DJ(\tau)$. These sums are easily extended to
partition functions for larger values of the central charge. In this
way, the
partition functions display a closer relation with the gravity
path integral. 

The appearance of the Farey transformed partition
function and why it is more reminiscent of the gravity path
integral remains mysterious (see also
\cite{Dijkgraaf:2000fq,Denef:2007vg}). A second puzzle is the
contribution of geometries without a proper classical realization to holomorphically factorizable
partition functions.

\section*{Acknowledgments}
I would like to thank Erik Verlinde for encouragement and stimulating
discussions. This research is supported by the Foundation of
Fundamental Research on Matter (FOM).

\end{document}